\documentclass[
 aip,
 apl,
 amsmath,amssymb,
 reprint,
]{revtex4-1}

\usepackage[]{graphicx}
\usepackage{dcolumn}
\usepackage{bm}
\usepackage[]{hyperref}
\hypersetup{
    colorlinks=true, 
    linkcolor=blue,
    citecolor=blue,
    filecolor=blue,
    urlcolor=blue,
    }
\usepackage[all]{hypcap}

\begin{document}

\title[]{A new transfer technique for high mobility graphene devices on commercially available hexagonal boron nitride}

\author{P. J. Zomer}
\email{p.j.zomer@rug.nl}

\author{S. P. Dash}
\altaffiliation[Present address]{ 
Quantum Device Physics Labaratory, Department of Microtechnology and Nanoscience - MC2, Chalmers University of Technology, G\"oteborg, Sweden}

\author{N. Tombros}

\author{B. J. van Wees}

\affiliation{Physics of Nanodevices, Zernike Institute for Advanced Materials, University of Groningen, Groningen, The Netherlands
}

\date{\today}

\begin{abstract}
We present electronic transport measurements of single- and bilayer graphene on commercially available hexagonal boron nitride. We extract mobilities as high as  125~000~cm$^{2}$V$^{-1}$s$^{-1}$ at room temperature and 275~000~cm$^{2}$V$^{-1}$s$^{-1}$  at 4.2~K. The excellent quality is supported by the early development of the $\nu$~=~1 quantum Hall plateau at a magnetic field of 5~T and temperature of 4.2~K. We also present a new and accurate transfer technique of graphene to hexagonal boron nitride crystals.
This technique is simple, fast and yields atomically flat graphene on boron nitride which is almost completely free of bubbles or wrinkles. The potential of commercially available boron nitride combined with our transfer technique makes high mobility graphene devices more accessible.
\end{abstract}

\pacs{72.80.Vp, 81.05.ue, 73.43.-f}

\keywords{graphene, boron nitride, high mobility}

\maketitle

The electronic quality of a graphene device strongly depends on the electronic interaction between the graphene layer and the substrate. 
The most commonly used substrate, silicon oxide (SiO$_{2}$), limits the mobility of charge carriers in devices due to its roughness~\cite{IshigamiAtomicStruc}, the presence of trapped charges~\cite{AndoImpScatt} and SiO$_{2}$ surface phonons~\cite{ChenSiO2}. 
A straightforward way in reducing the perturbations to graphene is to remove the substrate and suspend the flake. 
Mobilities as high as 230~000~cm$^{2}$V$^{-1}$s$^{-1}$ at charge carrier densities of n~=~2~$\times$~10$^{11}$cm$^{-2}$ and T~=~$\sim$~5~K have been obtained this way~\cite{BolotinSuspended,TombrosSuspendedPolymer}. 
However, due to mechanical instability under large applied gate voltages, it is not possible to induce high charge carrier densities in suspended devices. 
Furthermore, the lack of a supporting substrate strongly limits the fabrication of more complex graphene device geometries.

A recent breakthrough was the improvement in device electronic quality by using specially grown ultrapure hexagonal boron nitride (h-BN) crystals~\cite{TaniguchiHBN} as a substrate material~\cite{DeanBN}, reaching mobilities of 140~000~cm$^{2}$V$^{-1}$s$^{-1}$ near the charge neutrality point. 
Single h-BN crystals are atomically flat and contain a very low amount of charged impurities. 
STM studies confirm the considerable reduction of charge carrier inhomogeneity for graphene on h-BN compared to graphene on SiO$_2$ for both ultrapure~\cite{XueSTM} and commercially obtained h-BN~\cite{DeckerSTM}. 
Consequently the fractional quantum Hall effect~\cite{DeanMcQHE}, unconventional quantum Hall effect in tri-layer graphene~\cite{TaychatanapatTLGQHE} and ballistic transport at room temperature~\cite{MayorovBallisticRT} have been observed in BN-supported graphene devices. 

In this Letter we present electronic transport measurements in graphene on commercially available h-BN crystals (Momentive, Polartherm grade PT110). 
These crystals are of smaller size (on average $\sim$45~$\mu$m) than their high purity counterparts~\cite{TaniguchiHBN}.
With the use of a new and fast fabrication method containing only one cleaning step
we are nevertheless able to obtain mobilities as high as 120~000~cm$^{2}$V$^{-1}$s$^{-1}$ at room temperature and up to 275~000~ cm$^{2}$V$^{-1}$s$^{-1}$ at 4.2~K.
The excellent electronic quality is additionally confirmed by magnetotransport measurements.

\begin{figure}[b]
\includegraphics[width=85mm]{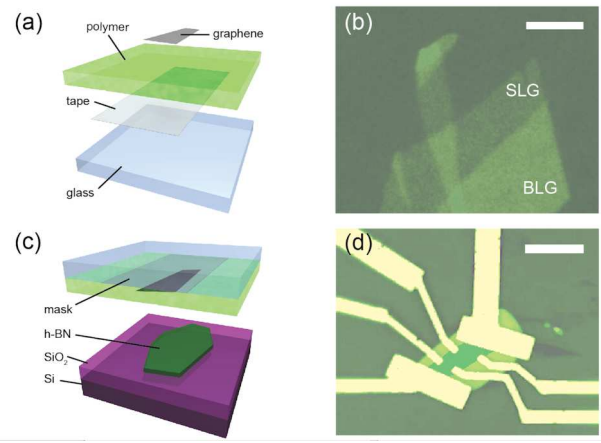}
\caption{
\label{fig:Fig1} (Color online) 
(a) Schematic of the transfer mask.
(b) SLG and BLG areas on the mask, the scale bar equals 5~$\mu$m.  
(c) Graphene alignment and transfer to a h-BN crystal. 
(d) Graphene flake in a Hall bar geometry on h-BN. The graphene and contacts are colored for clarity. The scale bar is 10~$\mu$m.
}
\end{figure} 

For the device fabrication we developed a simple and fast transfer technique. 
We first deposit thin (10-60~nm) h-BN crystals on a Si wafer with 500~nm SiO$_{2}$ by mechanical cleavage of h-BN powder using adhesive tape. 
By optical and atomic force microscopy (AFM) we select h-BN crystals larger than 10~$\mu$m in length for further processing. 
Next we prepare an optical mask for the transfer process.
The mask consists of 3 layers, as schematically depicted in Fig.\ref{fig:Fig1}a. 
The first layer is a $\sim$1.85~mm thick mechanically solid transparent glass. The second layer is an adhesive tape (Pritt) which is flexible and highly transparent, with the adhesive side facing the glass plate. 
The third and final layer is a spincoated $\sim$1~$\mu$m thick layer of methyl/n-butyl methacrylate copolymer (Elvacite 2550 acrylic resin) dissolved in methyl isobutyl ketone (MIBK), with a glass transition temperature of 36~$^{\circ}$C and a molecular weight of 98~K.
The mask is baked for 10 minutes at 120~$^{\circ}$C in order to remove the MIBK solvent from the copolymer.
Using standard mechanical exfoliation the graphene (SPI, HOPG grade ZYA) is deposited on the polymer layer.
Subsequent selection of graphene flakes is done by optical microscopy in the reflection mode using a CCD camera.  
The highest optical contrast is obtained by setting the CCD sensitivity to green light, for which we obtain a contrast of 6\% for single layer graphene and 12\% for bilayer graphene (Fig.\ref{fig:Fig1}b). 

In the following step we transfer a selected graphene flake onto an h-BN crystal. 
We modified a mask aligner (Karl Suss MJB-3) such that we can heat up the sample holder. 
During the transfer process we fix the h-BN containing substrate on the holder and set the temperature in the range between 75-100~$^{\circ}$C. 
Using the optical mask micromanipulators we align the graphene flake to the h-BN crystal and lower the polymer side of the mask onto the hot substrate, as shown in Fig.\ref{fig:Fig1}c. 
When the polymer touches the substrate it melts and makes strong contact with the SiO$_2$ surface. 
We then lift the mask and while the polymer is released from the adhesive tape layer on the mask, it remains attached to the sample. 

Subsequent to the transfer step the polymer is removed using acetone, followed by isopropanol rinsing. 
Remarkably, the majority of transferred graphene flake area (typical 95-100\% area for a total graphene area of 100~$\mu$m$^2$) is free of bubbles and wrinkles. 
This is a great improvement compared to other graphene transfer techniques which often result to a large amount of bubbles and wrinkles~\cite{MayorovBallisticRT}, being undesirable since they can reduce the electronic quality of graphene considerable and making additional etching a necessity. 
The amount of bubbles and wrinkles in graphene on h-BN probably depends on the initial strain between the graphene layer and the polymer substrate immediately before the transfer on h-BN. 
We believe that the strain between the graphene layer and the copolymer we use is much lower compared to the strain found between the ultra-thin ($<1~\mu$m) suspended PMMA layer and graphene in other transfer methods~\cite{DeanBN,MayorovBallisticRT}. 

After the transfer process we fabricate the electronic device using standard electron beam lithography and deposit titanium/gold (5~nm/75~nm) contact material using an e-gun evaporation system (Fig.\ref{fig:Fig1}d). 
The polymer mask used for processing is PMMA 950K, dissolved in ethyl-lactate. 
Lift off is done in acetone, followed by isopropanol rinsing.
Only after finalizing the device we anneal it once at 330~$^{\circ}$C in Ar/H$_2$ (85\%/15\%) flow for 8 hours in order to remove the polymer contaminants on the graphene surface. 

\begin{figure}[t]
\includegraphics[width=85mm]{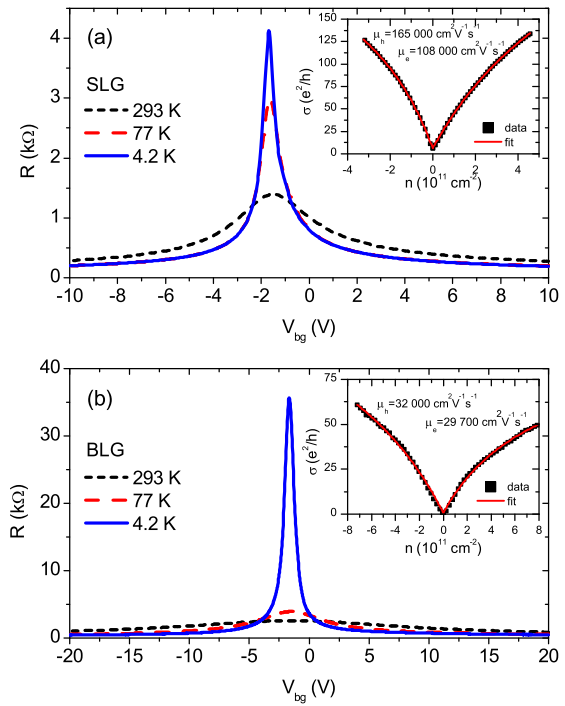}
\caption{
\label{fig:Fig2} (Color online) 
(a) Graphene resistivity as a function of the backgate voltage. The inset shows a fit of the conductivity at 4.2~K. 
(b) Bilayer graphene resistivity, the inset shows the fitted conductivity.
}
\end{figure} 

In order to extract the mobility from our devices we measured the dependence of the resistance on the backgate voltage applied to the Si wafer. 
We extract the gate induced charge carrier density in the graphene flake, $n=C_{g}/e~(V-V{_D})$, where $V_{D}$ is the voltage corresponding to the maximum resistivity (charge neutrality point) and the geometric capacitance $C_{g}$~$\approx$~6.3~nF for 500~nm thick SiO$_2$ and $\sim$50~nm thick h-BN. 
The mobility is determined as $\mu=1/ne\rho$, where $\rho$ is the resistivity of the flake. At the deflection point, which is defined as the charge carrier density for which the derivative d$\rho$/dn has an extreme, we find charge carrier mobilities as high as 275~000~cm$^{2}$V$^{-1}$s$^{-1}$ for n~$\approx$~0.8~$\times$~10$^{10}$cm$^{-2}$ (-1.9~V) at 4.2~K for the data presented in Fig.\ref{fig:Fig2}a. 
The room temperature mobility is found to be 125~000~cm$^{2}$V$^{-1}$s$^{-1}$ at n~$\approx$~4.3~$\times$~10$^{10}$cm$^{-2}$. 

Alternatively, for direct comparison with results obtained using ultrapure h-BN~\cite{DeanBN}, we fit the inverse conductivity using $1/\sigma=1/(ne\sigma+\sigma _{0})+\rho_s$ to obtain the mobility~\cite{Hwang2Dtransp}. 
Here, $\sigma _0$ is the conductivity at the charge neutrality point and $\rho_s$ is the contribution of short range scattering. 
This way we find for holes $\mu_h$~=~165~000~cm$^{2}$V$^{-1}$s$^{-1}$ and $\rho_{s,h}$~=~90~$\Omega$ and for electrons $\mu_e$~=~108~000~cm$^{2}$V$^{-1}$s$^{-1}$ and $\rho_{s,e}$~=~69~$\Omega$. 
Both methods for obtaining the mobility yield very high values, comparable to the best values reported so far for h-BN-based graphene devices~\cite{DeanBN,DeanMcQHE,MayorovBallisticRT}. 

In addition we measured one bilayer graphene sample (Fig.\ref{fig:Fig2}b). 
At the deflection point (n~$\approx$~1~$\times$~10$^{10}$cm$^{-2}$) we find for holes $\mu_h$~=~23~000~cm$^{2}$V$^{-1}$s$^{-1}$. 
Fitting the conductivity yields 32~000~cm$^{2}$V$^{-1}$s$^{-1}$.

In total we prepared 17 single layer graphene devices for which the Ar/H$_2$ cleaning step duration varied between 1 and 8 hours. 
All devices exhibited high mobilities. 
In order to investigate the reproducibility we prepared 9 devices in exactly the same way, following the recipe described in this Letter. 
The mobility at the deflection point was found to range from 80~000~cm$^{2}$V$^{-1}$s$^{-1}$  to 260~000~cm$^{2}$V$^{-1}$s$^{-1}$  with a median of 140~000~cm$^{2}$V$^{-1}$s$^{-1}$  at 77~K. Room temperature mobilities are found in the range from 55~000~cm$^{2}$V$^{-1}$s$^{-1}$  to 125~000~cm$^{2}$V$^{-1}$s$^{-1}$  with a median of 70~000~cm$^{2}$V$^{-1}$s$^{-1}$. 
Mobilities at high carrier densities ($\sim$7~$\times$~10$^{11}$cm$^{-2}$) lie between 23~000~cm$^{2}$V$^{-1}$s$^{-1}$  and 67~000~cm$^{2}$V$^{-1}$s$^{-1}$  for measurements at 77~K while room temperature mobilities  yield 18~000~cm$^{2}$V$^{-1}$s$^{-1}$  to 46~000~cm$^{2}$V$^{-1}$s$^{-1}$. 
These measurements indicate the good reproducibility of high mobility graphene devices based on commercially available h-BN using our fabrication recipe.

To further quantify the quality of our devices we look at the full width half maximum (FWHM) of the resistivity peak. 
This gives an indication about the electron-hole puddle induced charge carrier inhomogeneity. 
For the data in Fig.\ref{fig:Fig2}a we find a FWHM of $\sim$0.6 V at 4.2~K which confirms the low amount of charge carrier fluctuations in our device, being less than 2.2~$\times$~10$^{10}$cm$^{-2}$. 
This upper bound estimate for the inhomogeneity is slightly smaller than found for devices based on ultrapure h-BN, using the same analysis~\cite{DeanBN}.

\begin{figure}[t]
\includegraphics[width=85mm]{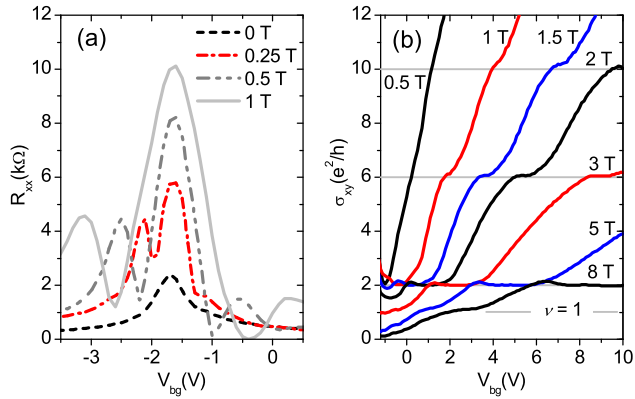}
\caption{
\label{fig:Fig3} (Color online) 
(a) R$_{xx}$  peaks indicate quantization setting in at 250~mT. 
(b) Hall conductivity of single layer graphene as function of the backgate voltage for magnetic fields ranging from 500~mT to 8~T. 
The measurements reveal the development of the $\nu$~=~1 plateau already at 5~T.
}
\end{figure} 

For additional confirmation of the high graphene device quality we conducted magnetotransport measurements up to 8~T at 4.2~K.
For this purpose we prepared several devices in the Hall bar geometry as shown in Fig.\ref{fig:Fig1}d. 
Typically we can see traces of quantization developing already at fields between 250 and 500~mT (Fig.\ref{fig:Fig3}a).
Another clear indication of the excellent quality of our devices is the development of the plateau at $\nu$~=~1 at magnetic fields as low as 5~T (Fig.\ref{fig:Fig3}b). 

In conclusion we show that graphene electronic devices we prepared on commercially obtained h-BN yield very high mobilities compared to previous reported graphene devices on highly purified h-BN crystals. 
We find mobilities up to 125~000~cm$^{2}$V$^{-1}$s$^{-1}$ at room temperature and 275~000~cm$^{2}$V$^{-1}$s$^{-1}$  at 4.2~K. 
This shows that commercially obtained h-BN, which is available in large amounts, offers a good alternative to non-commercial ultrapure h-BN.
Additionally, since the fabrication recipe we developed requires only one cleaning step it allows for fast device preparation of graphene on h-BN with almost no bubbles.
Given the current effort on high mobility graphene research, this work makes high quality graphene on h-BN broadly accessible.
\\  
    
We acknowledge B. Wolfs, J. G. Holstein and H. M. de Roosz for their technical assistance. 
This work is financially supported by the Dutch Foundation for Fundamental Research on Matter (FOM), NWO, NanoNed and the Zernike Institute for Advanced Materials.

\providecommand{\noopsort}[1]{}\providecommand{\singleletter}[1]{#1}%

\end{document}